\documentstyle[12pt]{article}
\newtheorem{theo}{Theorem}

\newtheorem{lemm}[theo]{Lemma}

\def\beq{\begin{equation}}
\def\eeq{\end{equation}}
\def\bea{\begin{eqnarray}}
\def\eea{\end{eqnarray}}
\def\beas{\begin{eqnarray*}}
\def\eeas{\end{eqnarray*}}
\def\nn{\nonumber}
\def\al{\alpha}
\def\be{\beta}

\def\la{\lambda}

\def\ps{\psi}
\def\Ga{\Gamma}
\def\Om{\Omega}
\def\De{\Delta}
\font\twelvemsbm=msbm10 at 12 true pt
\font\eightmsbm=msbm8
\font\sevenmsbm=msbm7
\newfam\msbmfam
\textfont\msbmfam=\twelvemsbm
\scriptfont\msbmfam=\eightmsbm
\scriptscriptfont\msbmfam=\sevenmsbm
\def\Bb#1{{\fam\msbmfam\relax#1}} 
\def\Nat{\Bb N}
\def\Zah{\Bb Z}
\def\Real{\Bb R}

\def\C{\Bb C}

\def\rn#1{\uppercase\expandafter{\romannumeral #1}}
\begin{document}
\addtolength{\baselineskip}{2mm}
\begin{center}
{\Large \bf Polynomial deformations of $osp(1/2)$ and
generalized parabosons}\\[2cm]
{\bf J.\ Van der Jeugt\footnote{Senior Research
Associate of N.F.W.O. (National Fund for Scientific Research of Belgium).}
and R.\ Jagannathan\footnote{Permanent address~: The Institute of Mathematical
Sciences, Madras 600 113, India.}}
\end{center}
\vskip 5mm
\noindent
Department of Applied Mathematics and Computer Science,
University of Ghent,\\ Krijgslaan 281-S9, B-9000 Gent, Belgium.\\[2mm]
E-mail : Joris.VanderJeugt@rug.ac.be and jagan@imsc.ernet.in.
\vskip 1cm
\noindent
Short title~: Generalized paraboson algebra.
\vskip 2cm
\noindent {\bf Abstract:}
We consider the algebra $R$ generated by three elements $A,B,H$
subject to three relations $[H,A]=A$, $[H,B]=-B$ and
$\{A,B\}=f(H)$. When $f(H)=H$ this coincides with the Lie
superalgebra $osp(1/2)$; when $f$ is a polynomial we speak of
polynomial deformations of $osp(1/2)$. Irreducible
representations of $R$ are described, and in the case
$\deg(f)\leq 2$ we obtain a complete classification, showing
some similarities but also some interesting differences with the
usual $osp(1/2)$ representations. The relation with deformed
oscillator algebras is discussed, leading to the interpretation
of $R$ as a generalized paraboson algebra.

\newpage

\section{Introduction}

Define the algebra $R=\C [A,B,H]$ subject to the following
relations~:
\beq
HA-AH=A,\qquad HB-BH=-B,\qquad AB+BA=f(H),
\label{defrel}
\eeq
where $f$ is a fixed polynomial function, $f\in \C[x]$. For many
of the applications that follow, $f$ can also be another
analytic function, but it will be clear that the polynomial case
itself is already a rich structure.
The purpose of the present paper is to study this algebra and in
particular its simple modules (irreducible representations). The
algebra $R$ is equal to the enveloping algebra of the Lie
superalgebra $osp(1/2)$ when $f(x)=x$, and we shall draw some
similarities and differences with the representation theory of
$osp(1/2)$ for the general case.  In this sense, the
algebra~(\ref{defrel}) can be considered as a polynomial
deformation of $osp(1/2)$. If the last relation
in~(\ref{defrel}) is replaced by $AB-BA=f(H)$, the
representation theory becomes rather different and in that case
one is dealing with polynomial deformations of the enveloping
algebra of $sl(2)$. These algebras have been studied by
Smith~\cite{smith} and were shown to have a surprisingly rich theory of
representations; from the physical point of view they are
non-canonical Heisenberg algebras, and have been investigated by
Brodimas {\em et al}~\cite{brodimas}.

\def\ad{a^\dagger}

There is clearly a relationship between the algebra $R$ and
generalized deformed oscillator
algebras~\cite{bonatsos,daskaloyannis,odaka,beckers,shanta,vanderjeugt}
generated by
the operators $a,\ad,N$ subject to the relations
\beq
[N,\ad]=\ad,\qquad [N,a]=-a,\qquad \ad a=\tilde F(N),\qquad
a\ad=\tilde F(N+1),
\label{defosc}
\eeq
where $\tilde F$ is called a structure function. If $\tilde
F(0)=\tilde F(p+1)=0$ and
$\tilde F(n)>0$ for $n\in\{ 1,2,\ldots, p\}$, this algebra is referred
to as a generalized parafermionic
algebra~\cite{quesne}. These algebras are related to nonlinear
deformations of $so(3)$ or $sl(2)$~\cite{quesne,polychronakos,rocek,pan}.
If $\tilde F(\la)=0$ and $\tilde F(\la+n)>0$ for
$n\in\{1, 2, \ldots\}$ this algebra can be referred to as a
generalized parabosonic algebra; another type of generalizations are
the $q$-deformed paraboson algebras which have been the topic of
many
papers~\cite{odaka,floreanini,celeghini,krishna,palev,chakrabarti,macfarlane}.
It should be mentioned
that for generalized deformed oscillator algebras there is also
the condition of unitarity of the
representations. When we study the simple modules of $R$, our
approach will be more general~: first we study all irreducible
representations of $R$, and only in a final stage we will
determine which of the representations are unitary. The
mathematical classification of the irreducible representations
(simple modules) is interesting on its own and deserves special
attention here.

When studying simple modules of $R$, we shall consider only the
highest weight modules. These are modules $V$ generated by a
vector $v$
satisfying $Av=0$ and $Hv=\la v$. Then $v$ is highest weight
vector of weight $\la$ ($\la\in\C$).

The structure of the paper is as follows. In section~{\rn 2} the
algebra $R$ is considered when $\deg(f)\leq 1$, and shown to be
familiar. In section~{\rn 3} we give a number of general properties of
$R$, including a differential realization and the structure of
its center. In the following section some general theory of
simple modules for $R$ is developed, and in section~{\rn 5} this is
illustrated for the case $\deg(f)=2$. The results in
section~{\rn 5}
are very surprising, and show how rich the representation theory
is. In section~{\rn 6} we give a number of comments on real and
unitary representations. Finally in section~{\rn 7} we indicate
the relation with generalized oscillator systems, and show how certain
one-dimensional quantum mechanical systems fall into the picture.

To conclude the present section, we fix some notation. For integers we use
the common symbol $\Zah=\{\ldots, -2 , -1 , 0 , 1 , 2 ,
\ldots\}$; for the nonnegative
integer numbers we use $\Nat=\{0,1,2,\ldots\}$; for the
positive integers we use $\Nat^*=\{1,2,\ldots\}$. We shall also
use the following type of notations~:
$2\Nat+1=\{1,3,5,\ldots\}$, $2\Nat^* =\{ 2,4,6,\ldots\}$, etc.

\def\deg{\hbox{deg}}
\section{The case $\deg(f)\leq 1$}

We have already noticed that when $f(x)=x$ the algebra $R$
coincides with the enveloping algebra of the Lie superalgebra
$osp(1/2)$. More generally, we shall see that when $\deg(f)\leq
1$, $R$ is familiar. We shall study the simple modules of $R$
when $\deg(f)\leq 1$.

Let $f(x)=\al x + \be$. There are three cases to be
distinguished.\\ [2mm]
Case (1)~: $\al=0,\; \be\ne 0$.\\
Let $V$ be a highest weight module with highest weight vector
$v_\la$, thus $Av_\la=0$ and $Hv_\la=\la v_\la$. Define
$v_{\la-1}= Bv_\la$; this vector is nonzero because otherwise
the relation $AB+BA=\be$ is not valid when acting on $v_\la$.
Moreover, it follows from the defining relations that
$Hv_{\la-1}= (\la-1)v_{\la-1}$ and $Av_{\la-1}=\be v_\la$. Next,
consider $v_{\la-2}=Bv_{\la-1}$. One finds that $Av_{\la-2}=
ABv_{\la-1}= (\be-BA)v_{\la-1}=0$. Hence if $v_{\la-2}$ were
nonzero, it would be an element of a nonzero submodule of $V$.
But since $V$ is simple, this submodule must be zero and
$v_{\la-2}=0$. Consequently, all simple modules of $R$ are
two-dimensional, and their explicit matrix representations are
given by~:
\beq
A=\left(\begin{array}{cc} 0&\be\\0&0\end{array}\right),\qquad
B=\left(\begin{array}{cc} 0&0\\1&0\end{array}\right),\qquad
H=\left(\begin{array}{cc} \la&0\\0&\la-1\end{array}\right).
\eeq
\vskip 2mm
Case (2)~; $\al=\be=0$.\\
Again, let $v_\la$ be the highest weight vector of a simple
module $V$, thus $Av_\la=0$ and $Hv_\la=\la v_\la$. Consider
$v_{\la-1} = B v_{\la}$; from $AB+BA=0$, it follows that
$Av_{\la-1}=0$, thus $v_{\la-1}$ belongs to a submodule. Since
$V$ is simple, one concludes that $v_{\la-1}=0$. Thus all simple
modules are one-dimensional, and are given by~:
\beq
Av_\la = B v_\la =0,\qquad Hv_\la=\la v_\la.
\eeq
\vskip 2mm
Case (3)~: $\al\ne 0$.\\
Let
\beq
\tilde H=H+\be/\al,\qquad \tilde A=A/\al,\qquad\hbox{and}\qquad\tilde B=B.
\label{tilde}
\eeq
The
relations~(\ref{defrel}) with $f(x)=\al x + \be$ become
\beq
\tilde H\tilde A-\tilde A\tilde H = \tilde A,\qquad
\tilde H\tilde B-\tilde B\tilde H = -\tilde B,\qquad
\tilde A\tilde B+\tilde B\tilde A = \tilde H.
\label{tildeR}
\eeq
These are the defining relations of the enveloping algebra of
$osp(1/2)$. But due to the linear relationship
there is a one-to-one correspondence
between the simple modules of $R$ and the simple modules of the
algebra generated by $\tilde A, \tilde B$ and $\tilde H$,
subject to (\ref{tildeR}).  Since all simple modules of
$osp(1/2)$ are known~\cite{scheunert,hughes},
all simple modules of $R$ follow directly.
In particular, for every $j\in\Nat$, there exists (up to
isomorphism) one simple module of $R$ with dimension $2j+1$; and
there exist no simple modules with even dimension.

\section{Some structure theorems for the general case}

First of all, let us give some simple realizations of $R$, with
$f$ a polynomial of degree $>1$. For $m,n\in\Nat$, define
the following polynomial~:
\beq
p_{m,n}(t)= t\prod_{j=1}^m (t-j)\prod_{k=1}^n (t-k).
\eeq

\begin{lemm}
With $x$ as multiplication by $x$ and $\partial$ as
differentiation with respect to $x$ on $\C[x]$, the realizations
\beq
A=x^{n+1}\partial^n,\qquad B=x^m\partial^{m+1},\qquad
H=x\partial \qquad (m,n\in\Nat),
\eeq
satisfy the relations~(\ref{defrel}) with
$f(t)=p_{m,n}(t)+p_{m,n}(t+1)$.
\end{lemm}
{\em Proof.} First one shows that $AB=p_{m,n}(H)$. Since
\beq
x\partial(x^{n+1}\partial^n)=(n+1)x^{n+1}\partial^n+x^{n+2}\partial^{n+1},
\eeq
we have that $x^{n+2}\partial^{n+1}=(H-n-1)x^{n+1}\partial^n$,
and thus
\beq
x^{n+2}\partial^{n+1}x^m\partial^{m+1}=
(H-n-1)x^{n+1}\partial^{n}x^m\partial^{m+1}.
\eeq
Similarly, one finds
\beq
x^{n+1}\partial^{n}x^{m+1}\partial^{m+2}=
x^{n+1}\partial^{n}x^m\partial^{m+1}(H-m-1).
\eeq
Then $AB=p_{m,n}(H)$ follows by induction on $n$ and $m$. The
relation $BA=p_{m,n}(H+1)$ is proved similarly, or follows from
the general observation that $BF(H)=F(H+1)B$.\hfill$\Box$

Next, we consider some general structure results.
The three relations~(\ref{defrel}) allow one to reorder any
polynomial in $A,B$ and $H$ unambiguously. Hence we have
\begin{theo}
A basis for $R$ is given by the monomials $A^iB^jH^k$,
$(i,j,k\in\Nat)$.
\label{pbw}
\end{theo}

It is easy to see that $[H,A^iB^jH^k]=(i-j)A^iB^jH^k$. Thus we
have a weight space decomposition for $R$~:
$R=\oplus_{\mu\in\Zah} R_\mu$, where $R_\mu=\{x\in R \;|\;
[H,x]=\mu x\}$. From the defining relations
one can deduce that $R_0=\C[AB,H]$; in other words, the
commutant of $H$ consists of polynomials in $AB$ and in $H$.

For the next theorem, we need an easy lemma.
\begin{lemm}
Let $f\in\C[x]$.
There exists a unique polynomial $F\in \C[x]$ such that $F(x+1)+F(x)=f(x)$.
\label{lem2}
\end{lemm}
{\em Proof.}
Assume that $\deg(f)=n$. Since
$\deg(F(x+1)+F(x)) = \deg(F)$, let $F(x)=\sum_{k=0}^n a_k
x^k$. Then
$$
F(x+1)+F(x)=\sum_{k=0}^n \left( \sum_{j=k}^n a_j\left(j\atop
k\right) + a_k\right) x^k.
$$
Now $F(x+1)+F(x)=f(x)$ reduces to a nonsingular upper
triangular system in the unknowns $a_0, a_1,\ldots ,a_n$, yielding a unique
solution for $F$.\hfill $\Box$

Consider the algebra $R$ with relations (\ref{defrel})
characterized by $f$, and $F$ determined by the previous lemma.

\begin{lemm}
The element $\Om=A^2B^2+AB\left(F(H)-F(H-1)\right)-F(H)^2$
is a central element of $R$.
\label{lemma}
\end{lemm}
{\em Proof.}
By straightforward calculation one finds that
\beas
[A,A^2B^2]&=&-A^2Bf(H)+A^2B f(H-1),\\[0pt]
[A,ABF(H)]&=&
A^2BF(H)-Af(H)F(H+1) +A^2B F(H+1)\\
 &=& A^2Bf(H)-Af(H)F(H+1),\\[0pt]
[A,-ABF(H-1)]&=& -A^2Bf(H-1)+Af(H)F(H),\\[0pt]
[A,-F(H)^2]&=&-AF(H)^2+AF(H+1)^2.
\eeas
Thus $[A,\Om]=0$;
similarly one finds that $[B,\Om]=0$, and with $[H,\Om]=0$ it
follows that $\Om$ is central.\hfill$\Box$

In fact, every other central element of $R$ is a polynomial expression
in $\Om$~:
\begin{theo}
The center of $R$ is equal to $\C [\Om]$.
\end{theo}
{\em Proof.}
Let $z$ be a central element; since $z$ commutes with $H$ it
belongs to $R_0$, and we have seen that $R_0=\C [AB,H]$. But
$(AB)^2=-A^2B^2+ABf(H-1)$, thus
\beq
R_0 = \C [\Om,H]\oplus AB\;\C [\Om,H].
\eeq
Therefore, $z$ can be written as
\beq
z=\sum_{k=0}^m \Om^k c_k + AB\sum_{k=0}^m\Om^k d_k,\qquad c_k,d_k\in
\C[H].
\eeq
{}From theorem~\ref{pbw} and the explicit expression of $\Om$, it
follows that
\beq
\Om^k=\sum_{i=0}^{2k} A^iB^i g^{2k}_i,
\label{om}
\eeq
where $g^{2k}_i$ is a polynomial in $H$ and $g^{2k}_{2k}=1$.
Next, we express that $z$ is central~:
\bea
0\equiv [A,z]&=& \sum_{k=0}^m
A\Om^k\left(c_k(H)-c_k(H+1)\right)+ \nn \\
& & \sum_{k=0}^m 2A^2B\Om^k\left(d_k(H)+d_k(H+1)\right) +
\sum_{k=0}^m A\Om^k f(H)d_k(H+1).
\label{Az}
\eea
Rearranging all terms of~(\ref{Az}) according to the basis of
theorem~\ref{pbw} using (\ref{om}), and considered as a
polynomial in $A$ and $B$
with coefficients in $\C[H]$, the term of highest degree is
$A^{2m+2}B^{2m+1}\left( d_m(H)+d_m(H+1)\right)$. Clearly this
term has to vanish (theorem~\ref{pbw}), and this can happen only
when $d_m=0$. Using the fact that $d_m=0$, the remaining term of
highest degree in (\ref{Az}) is then
$A^{2m+1}B^{2m}\left(c_m(H)- c_m(H+1) \right)$. Again, this has
to vanish, implying that $c_m$ is just a constant~: $c_m\in
\C$. But now we can apply the above argument to $z-c_m\Om^m$,
implying that $d_{m-1}=0$ and $c_{m-1}\in\C $, and so on. So
finally all $d_k$ are zero and all $c_k$ constants.\hfill$\Box$

The next result gives the eigenvalue of $\Om$ acting on a
highest weight module.

\begin{theo}
Let $V$ be an $R$ module generated by a highest weight vector
$v_\la$. Then, for all $v\in V$,
\beq
\Om v = -F(\la+1)^2 v,
\label{omeigval}
\eeq
where $F$ is defined by means of Lemma~\ref{lem2}.
\label{omeig}
\end{theo}
{\em Proof.}
The expression of $\Om$ given in Lemma~\ref{lemma} can be
rewritten in the following form~:
\beq
\Om=AB^2A+BA\left(F(H+1)-F(H)\right)-F(H+1)^2.
\eeq
Thus
\beq
\Om v_\la = -F(\la+1)^2 v_\la,
\eeq
and since the module is generated by $v_\la$, $\Om$ has the same
eigenvalue for every $v$ in $V$. \hfill$\Box$

\section{Simple modules of $R$}

In this section we shall study some general features of simple
modules of $R$, where $f(x)$ is a general polynomial of degree $n$. The {\em
Verma module} $V(\la)$ can be identified with the vector space
with basis
\beq
v_{\la-j} = B^j v_\la,\qquad j\in\Nat,
\eeq
where $v_\la$ is a highest weight vector satisfying $A v_\la =0$
and $H v_\la=\la v_\la$. The Verma module $V(\la)$ contains a
maximal submodule $M$; if $M=\{ 0\}$ then $V(\la)$ is simple,
otherwise the quotient module $L(\la)=V(\la)/M$ is a simple
module of highest weight $\la$. All simple highest weight
modules are obtained this way.

To see whether a vector $v_{\la-j}$ is in a submodule one
determines $A v_{\la-j}$. By induction, it is easy to prove that
\beq
AB^jv_\la = \left(f(\la-j+1)-f(\la-j+2)+f(\la-j+3)-\cdots
-(-1)^j f(\la)\right) B^{j-1} v_\la.
\label{avl-j}
\eeq
If $j$ is even, the coefficient in the rhs of (\ref{avl-j}) is
equal to
\beq
F(\la+1-j)-F(\la+1).
\label{eq0}
\eeq
Similarly, using Lemma~\ref{lem2}, one finds that for the case $j$
odd, the rhs of (\ref{avl-j}) is equal to
\beq
F(\la+1)-F(\la+2-j)+f(\la+1-j).
\label{eq1}
\eeq
For $\la\in\C$, we define~:
\bea
S_0(\la)&=&\{ j\in 2\Nat^* \;|\; F(\la+1)-F(\la+1-j)=0\} , \label{s0}\\
S_1(\la)&=&\{ j\in (2\Nat+1) \;|\;
F(\la+1)-F(\la+2-j)+f(\la+1-j)=0\}, \label{s1} \\
S(\la)&=&S_0(\la)\cup S_1(\la). \label{s}
\eea
Note that (\ref{eq0}) is of degree $n-1$ in
$\la$, and that (\ref{eq1}) is of degree $n$ in $\la$. As
equations in $j$, (\ref{eq0}) has the trivial solution $j=0$ (to
be excluded from $S_0(\la)$) so that an equation of degree $n-1$
remains, and (\ref{eq1}) is of degree $n$. Of course, for given
$\la$ not all solutions for these equations in $j$ will yield
integer $j$ so that $S_0(\la)$ and/or $S_1(\la)$ are often
empty.

The primitive vectors of $V(\la)$ are those $v_{\la-j}$ such
that $Av_{\la-j}=0$; they generate the submodules of $V(\la)$.
With the given definitions we have the following lemmas~:

\begin{lemm}
The submodules of $V(\la)$ are $\C[B]B^jV(\la)$ with
$j\in S(\la)$.
\end{lemm}

\begin{lemm}
Let $\la\in\C$.
\begin{itemize}
\item[(1)] If $S(\la)\ne\emptyset$ then $L(\la)=
V(\la)/B^j V(\la)$, where $j=\min S(\la)$, otherwise
$L(\la)=V(\la)$.
\item[(2)] For even $j$ ($j\in 2\Nat^*$) the number of simple modules
of dimension $j$ is equal to $\#\{\la\in\C \;|\;
F(\la+1)-F(\la+1-j)=0 \hbox{ and } j=\min S(\la)\}$.
\item[(3)] For odd $j$ ($j\in 2\Nat+1$) the number of simple modules
of dimension $j$ is equal to $\#\{\la\in\C \;|\;
F(\la+1)-F(\la+2-j)+f(\al+1-j)=0 \hbox{ and } j=\min S(\la)\}$.
\end{itemize}
\end{lemm}
Because of the previous remarks concerning the degrees of the
relevant equations, it follows that for even $j$ there are at
most $n-1$ simple modules of dimension $j$ and that for odd $j$
there are at most $n$ simple modules of dimension $j$.
Generally there will also be $n-1$ simple modules for even
dimensions and $n$ modules for odd dimensions, apart from
certain exceptions. This will be illustrated in the following
section, where a complete study of $\deg(f)=n=2$ is presented.

\section{The case $\deg(f)=2$}

In the third relation of (\ref{defrel}) $f(H)$ is of the form
$aH^2+bH+c$ with $a\ne 0$. But due to the fact that the representation
theory of $R$ is equivalent to the representation theory of the
algebra generated by (\ref{tilde}) (subject to the corresponding
relations), there are in fact two degrees of freedom that one can delete.
Therefore we shall assume in the rest of this section that the
algebra $R$ is determined by the following relations~:
\beq
HA-AH=A,\qquad HB-BH=-B,\qquad AB+BA=2H^2+2H+1-c,
\label{deg2}
\eeq
where $c\in\C$. The form of $f(H)$ in (\ref{deg2}) is somewhat
arbitrary, but for our purposes the one chosen here seems to be
the most appropriate. We emphasize again that for every given
$f$ of degree $2$ the algebra can be ``rescaled'' by means of a
transformation of type (\ref{tilde}) such that it becomes
(\ref{deg2}).

With $f(x)=2x^2+2x+1-c$, we have that
$F(x)=x^2-c/2$. Then the equations (\ref{eq0}) and
(\ref{eq1}) are easy to calculate and one finds respectively
$j(j-2\la-2)=0$ and $(j-2\la-2)^2=2c-j^2$. This implies the
following~:
\beq
\hbox{If }\la\in\Nat\hbox{ then } S_0(\la)=\{2\la+2\}
\hbox{ else } S_0(\la)=\emptyset,
\label{s02}
\eeq
\beq
S_1(\la)=\{ j\in(2\Nat+1) \hbox{ such that }
(j-\la-1)^2=c-(\la+1)^2\}.
\eeq

The purpose is now twofold~: for given $\la\in\C$, we wish to
give (in terms of $c$) the dimension of $L(\la)$
(theorem~\ref{la}); and for given
$j$, we wish to determine the number of (inequivalent) simple
modules of dimension $j$ (theorem~\ref{j}).

For the rest of this section it will be useful to work with
$\la+1$ instead of $\la$, and we even introduce a notation for
it~:
\beq
l=\la+1.
\eeq
Suppose $\la\in\C$ is given.\\
Consider first the case that
$\la\not\in\Nat$, then $S_0(\la)=\emptyset$. The equation
determining $S_1(\la)$ is given by
\beq
(j-l)^2=c-l^2.
\label{eqjl}
\eeq
Can we have
two integer solutions $j_1$ and $j_2$, both odd, for this
equation? From the equation it would then follow that
$j_1+j_2=2l$, which would imply that $l$ or $\la$ is in $\Nat$,
which is not the case. Hence, we can have at most one integer
solution for the equation (\ref{eqjl}). In order to have an
integer and odd solution for (\ref{eqjl}) $l\pm\sqrt{c-l^2}$
must be an element of $2\Nat+1$. This can only happen when $c$
is of the form $2l^2-2lm+m^2$ with $m\in (2\Nat+1)$. If $c$ is
of this form then $L(\la)$ is finite-dimensional with dimension
$m$, otherwise $L(\la)$ is infinite-dimensional.\\
Consider next the case that $\la\in\Nat$. Then
$S_0(\la)=\{2l\}$. Now we must examine whether $S_1(\la)$
contains odd integers less than $2l$. In order to have integer
solutions for $j$, (\ref{eqjl}) implies that $c-l^2$ should be a
square, or $c\in \{l^2+m^2 \;|\; m\in\Nat\}$. In that case the
solutions to (\ref{eqjl}) are $j=l\pm m$. Suppose first that $l$
is even, and $c=l^2+m^2$. If $m$ is also even then
$S_1(\la)=\emptyset$ and the dimension is given by $2l$. If $m$ is
odd we have that $S_1(\la)=\{l+m\}$ or $S_1(\la)=\{l+m,l-m\}$ (if
$m<l$). So if $m>l$ then the dimension is $2l$ anyway, otherwise
the dimension is $l-m$. Suppose next that $l$ is odd, and again
$c=l^2+m^2$. If $m$ is also odd, $S_1(\la)=\emptyset$ and the
dimension is $2l$. If $m$ is even, then, as in the previous
case, the dimension is $2l$ if $m>l$ and $l-m$ is $m<l$. The
conclusion is~:

\begin{theo}
Given $\la\in\C$ and $l=\la+1$.
\begin{itemize}
\item[(a)] If $\la\not\in\Nat$, let
$\Ga_\la=\{2l^2-2lm+m^2\;|\;m\in(2\Nat+1)\}$. If $c\not\in \Ga_\la$ then
$\dim L(\la)=\infty$; if $c\in \Ga_\la$ then $\dim L(\la)=m$
(where $m$ is the unique odd solution to $c=2l^2-2lm+m^2$).
\item[(b)] If $\la\in\Nat$, let $\De_\la=\{l^2+m^2\;|\; m\in\Nat\}$. If
$c\not\in \De_\la$ then $\dim L(\la)=2l$. If $c\in \De_\la$ and $c$
is even then $\dim L(\la)=2l$; if $c\in \De_\la$ and $c$ is odd
then $\dim L(\la)=2l$ when $c\geq 2l^2$ and $\dim
L(\la)=l-\sqrt{c-l^2}$ when $c< 2l^2$.
\end{itemize}
\label{la}
\end{theo}

The opposite question is even more interesting~: for every given
integer $j$, determine the number of simple modules of dimension
$j$. \\
Consider first the case that $j$ is even~: $j\in 2\Nat^*$. If we
put $\la=j/2-1$ then $S_0(\la)=\{j\}$. To see whether for this
$\la$ also $\dim L(\la)=j$, we have to examine $S_1(\la)$. But
according to the previous theorem, the dimension of $L(\la)$ is
less than $j$ only if $c<2l^2$ with $c=l^2+m^2$ odd and
$m\in\Nat$. It is appropriate to introduce the following set~:
\beq
I=\{ p^2+q^2 \;|\; p\in 2\Nat,\quad q\in 2\Nat+1\}.
\label{I}
\eeq
If $c\not\in I$ then there is one simple module of dimension $j$,
$j\in 2\Nat^*$. If $c\in I$, we define~:
\beq
K_c=\{ (k,m) \;|\; c=k^2+m^2\hbox{ with } m<k\hbox{ and
}k,m\in\Nat\}.
\label{Kc}
\eeq
If $c\in I$ then for every even $j$ there is one module of
dimension $j$, except when $j$ is of the form $2k$ with $(k,m)$
in $K_c$~: in this case $L(\la)$ has dimension $k-m<j=2k$ and
there is no module of dimension $j$.\\
Consider next the case that $j$ is odd. From the discriminant of
(\ref{eqjl}) one can see that this has two distinct solutions for
$l$ if $c-j^2/2\ne 0$, or if $c\not\in J$ where
\beq
J=\{ k^2/2 \;|\; k\in 2\Nat+1\}.
\label{J}
\eeq
To see when the solutions for $\l$ are integers, rewrite
equation (\ref{eqjl}) as $2c=j^2+(2l-j)^2$. So this can have an
integer solution for $l$ only if $2c$ is of the form $u^2+v^2$,
with both $u$ and $v$ odd integers. So, let
\beq
I'=\{ u^2+v^2 \;|\; u,v\in (2\Nat+1)\}.
\eeq
If $2c\not\in I'$ then the solutions $l_1$ and $l_2$ for $l$ are
nonintegers and thus there are two modules of dimension $j$. If
$2c\in I'$, write $2c=u^2+v^2$ with $u<v$ (the case $u=v$
implies $c\in J$) and $u$ and $v$ odd. For $j=u$ the
solutions for $l$ are $(u+v)/2$ and $(u-v)/2$, and one can
verify that both of these yield a simple module of dimension
$j$; for $j=v$ the solutions for $l$ are $l_1=(u+v)/2$ and $l_2=(v-u)/2$,
and for both of these $\min S(\la)=u<j$ so there are no modules
of dimension $j$. A final step to summarize the situation is
given by

\begin{lemm}
$c\in I$ if and only if $2c\in I'$.
\end{lemm}
{\em Proof.}
The proof is rather trivial. If $c\in I$ then
$c=(2r)^2+(2s+1)^2$ with $r,s\in\Nat$. But then $2c=(2r+2s+1)^2+
(2r-2s-1)^2$, so $c\in I'$. Conversely, if $2c\in I'$, then
$2c=(2r+1)^2 + (2s+1)^2$ ($r,s\in\Nat$), and then
$c=(r+s+1)^2+(r-s)^2$. Since $r+s+1$ and $r-s$ have different
parity it follows that $c\in I$.\hfill$\Box$

We can now state the result.

\begin{theo}
Let $c\in\C$ and $I,J,K_c$ be defined by
(\ref{I}), (\ref{J}), (\ref{Kc}) respectively.
\begin{itemize}
\item[(a)] When $c\not\in I$ and $c\not\in J$ then for every even
dimension $j$ there is one simple module $V(\la)$ {\rm [}determined by
$l=\la+1=j/2${\rm ]} and for every odd dimension $j$ there
are two simple modules $V(\la_1)$ and $V(\la_2)$ {\rm [}determined by
the solutions $l_1$ and $l_2$ of $2l^2-2jl+j^2-c=0${\rm ]}.
\item[(b)] When $c\in J$, i.e.\ $c=k^2/2$ with $k\in(2\Nat+1)$, then
for every even dimension $j$ there is one simple module
{\rm [}determined by $l=j/2${\rm ]}, and for every odd dimension there are
two simple modules {\rm [}again determined by the solutions of
(\ref{eqjl}){\rm ]} except when $j=k$~: then there is only one simple
module {\rm [}determined by $l=k/2${\rm ]}.
\item[(c)] When $c\in I$ then for every even dimension $j$ there is one
simple module {\rm [}determined by $l=j/2${\rm ]} except for those $j$ of
the form $j=2k$ with $(k,m)\in K_c$; and for every odd dimension
$j$ there are two simple modules {\rm [}determined by (\ref{eqjl}){\rm ]}
except when $j$ is of the form $j=k+m$ with $(k,m)\in K_c$, in
which case there are no simple modules.
\end{itemize}
\label{j}
\end{theo}

To illustrate the theorem, consider two examples.\\
When $c=9/2$,
$R$ has one simple module in every even dimension, two simple
modules in every odd dimension different from 3, and one simple
module in dimension~3.\\
When $c=1105$, we see that $c\in I$ and in fact
\beq
K_{1105}=\{(33,4), (32,9), (31,12),(24,23)\}.
\label{list}
\eeq
Then $R$ has no modules of dimension 37, 41, 43, 47, 48, 62, 64
and 66. For the remaining even dimensions there is one simple
module, and for the remaining odd dimensions there are two
simple modules.

All this illustrates that the structure of simple modules of $R$
is very rich when $\deg(f)>1$.

We conclude this section with a remark. When $v_{\la-j}$ is a
primitive vector in $V(\la)$, it follows from the proof of
theorem~\ref{omeig} that $\Om v_{\la-j}=-F(\la-j+1)^2v_{\la-j}$.
This implies the relation $F(\la+1)^2=F(\la-j+1)^2$. On the other
hand when $F(\la+1)^2=F(\la-j+1)^2$ for some integer $j$, it does
not necessarily imply that $v_{\la-j}$ is a primitive vector. An
example is given by $c=9/2$ and $\la=3/2$; then
$F(5/2)=F(-5/2)$, but $v_{\la-5}$ is not primitive in $V(5/2)$. In
fact, $V(5/2)$ has no primitive vectors and is simple.

\section{Real and unitary representations}

In the previous sections we considered complex representations,
and in this section we shall consider $f\in\Real[x]$ and
examine the real simple modules of $R$. This is actually rather
easy~: one simply takes the classification of complex simple
modules of $R$, and keeps only those with real matrix elements
of $A$, $B$ and $H$. For the modules $L(\la)$, the matrix
elements of $A$ and $B$ are real anyway (if $f\in\Real[x]$), so
the only remaining condition is that also $\la$ must be real.

To illustrate this, consider again the case $\deg(f)=2$, and
suppose $R$ is now defined by (\ref{deg2}) with $c\in\Real$.
Then the results of theorem~\ref{j} yield also the real
representations of $R$, as long as the roots to be taken are
real, in other words (in the following module stands for real module)~:
\begin{itemize}
\item[(a)] When $c\not\in I$ and $c\not\in J$ then for every even
dimension $j$ there is one simple module $V(\la)$
and for every odd dimension $j<\sqrt{2c}$ there
are two simple modules; for odd dimensions $j>\sqrt{2c}$ there
are no simple modules.
\item[(b)] When $c\in J$, i.e.\ $c=k^2/2$ with $k\in(2\Nat+1)$, then
for every even dimension $j$ there is one simple module.
For every odd dimension $j<k$ there are
two simple modules; for $j=k$ there is one simple module; for
$j>k$ there are no simple modules.
\item[(c)] When $c\in I$ then for every even dimension $j$ there is one
simple module except for those $j$ of
the form $j=2k$ with $(k,m)\in K_c$. For every odd dimension
$j<\sqrt{2c}$ there are two simple modules except when $j$ is of
the form $j=k+m$ with $(k,m)\in K_c$ (in which case there are no
simple modules); for $j>\sqrt{2c}$ there are no simple modules.
\end{itemize}

We can consider the same example $c=1105$ as in the previous
section. So $R$ has one simple real module in every even
dimension different from 48, 62, 64 and 66. For the
odd-dimensional modules, note that $\sqrt{2c}\approx 47.0106$.
Thus, using also the results of (\ref{list}), $R$ has two
simple real modules in dimensions $1,3,5,\ldots,33,35$ and also in 39
and 45. In the remaining odd dimensions, $R$ has no simple real modules.

In order to define ``unitary representations'' for $R$, one
first considers a Hermitian operation on $R$. This is an
operation $*:R\rightarrow R$ which satisfies $(rs)^*=s^*r^*$,
$(\la r)^*=\bar\la r^*$ and $(r^*)^*=r$, for all $r,s\in R$ and
$\la\in\C$ (with $\bar\la$ the complex conjugate). The Hermitian
operation introduced here is as follows~:
\beq
A^*=B,\qquad B^*=A,\qquad H^*=H.
\eeq
This induces indeed a Hermitian operation on the whole of $R$
provided it leaves the defining relations invariant, that is
provided $f$ is a {\em real} polynomial.

Such a Hermitian operation $*$ induces a Hermitian form
$\langle\;|\;\rangle$ on the $R$ modules $L(\la)$ by~:
\bea
&&\langle v_\la | v_\la \rangle =1,\\
&&\langle r v | s w \rangle = \langle v | r^*s w\rangle,\qquad
r,s\in R,\qquad v,w\in L(\la).
\eea
In particular, such a form is non-degenerate on $L(\la)$ and
vectors of different weights are orthogonal with respect to this
form. If this form is positive definite, then it yields an inner
product on $L(\la)$, and then the representation is called {\em
unitary}.

Consider a module $L(\la)$ with highest weight vector
$v_\la$. In order to see when $L(\la)$ is unitary, we use
(\ref{avl-j})~:
\beq
\langle v_{\la-j}|v_{\la-j}\rangle =
\left(f(\la-j+1)-f(\la-j+2)+\cdots
-(-1)^j f(\la)\right) \langle v_{\la-j+1}|v_{\la-j+1}\rangle.
\label{matel}
\eeq
So, if $\dim L(\la) \geq 1$, the first condition is
$f(\la)>0$. If the dimension of $L(\la)$ is larger than 2,
the next matrix element yields
$f(\la-1)-f(\la)>0$. In the most general case one
continues like this and requires that all matrix elements in
(\ref{matel}) must be positive.

Here we shall again examine the case $\deg(f)=2$, and $R$
defined by (\ref{deg2}) with $c$ real. The condition
$f(\la-1)-f(\la)>0$ reduces to $\la<0$, and then automatically
the remaining matrix elements in (\ref{matel}) are also positive when
$f(\la)>0$, since~:
\bea
&&\langle v_{\la-j}|v_{\la-j}\rangle = j(j-2\la-2) \qquad
 \hbox{ for } j\in 2\Nat^*,\\
&&\langle v_{\la-j}|v_{\la-j}\rangle = (j-1)(j-2\la-1)+f(\la) \qquad
 \hbox{ for } j\in 2\Nat+1.
\eea
Thus, for the degree 2 case there are two
types of unitary representations~: if $L(\la)$ is 2-dimensional
then the only condition is $f(\la)>0$; else there are two
conditions~: $f(\la)>0$ and $\la<0$.

Consider first the 2-dimensional (real simple) modules. There is
at most one such module, namely $L(0)$. From theorem~\ref{la} we
deduce that $L(0)$ is 2-dimensional unless $c=1$. On the other
hand, $f(\la)>0$ implies $c<1$. The conclusion is as follows~:
for $c<1$ there is precisely one 2-dimensional unitary simple
module of $R$; the explicit matrix elements are in fact given
by~:
\beq
A=\left(\begin{array}{cc} 0&1-c\\0&0\end{array}\right),\qquad
B=\left(\begin{array}{cc} 0&0\\1&0\end{array}\right),\qquad
H=\left(\begin{array}{cc} 0&0\\0&-1\end{array}\right),
\label{2dim}
\eeq
with $\langle v_0|v_0\rangle=1$ and $\langle
v_{-1}|v_{-1}\rangle=1-c$ (note that by going to an orthonormal
basis the nonzero matrix elements of $A$ and $B$ would become
$\sqrt{1-c}$ and in matrix notation they would indeed be each
others Hermitian conjugate).

Consider next the remaining modules $L(\la)$ that must satisfy
both the conditions $f(\la)>0$ and $\la<0$. From
theorem~\ref{la}(a) one deduces that there are no such
finite-dimensional modules but only infinite-dimensional modules
$L(\la)$. Then, by examining the roots of the equation
$2\la^2+2\la+1-c=0$, one finds~:
\begin{itemize}
\item if $c<1/2$, the unitary simple modules are
$L(\la)=V(\la)$ with $\la<0$;
\item if $1/2\leq c\leq 5/8$, the unitary simple modules are
$L(\la)=V(\la)$ with either $\la<-1/2-\sqrt{2c-1}$ or else
$-1/2+\sqrt{2c-1}<\la<0$.
\item if $c>5/8$ the unitary simple modules are
$L(\la)=V(\la)$ with  $\la<-1/2-\sqrt{2c-1}$.
\end{itemize}
Of course, apart from the ones given above (which are all
infinite-dimensional), there is also the unique
finite-dimensional unitary real simple module $L(0)$ when
$c<1$.

This leads to an interesting remark. For the undeformed
paraboson algebra with $f(x)=x$, one is referred to the
representations of $osp(1/2)$, and thus
there are no finite-dimensional unitary representations; the
only unitary representations are infinite-dimensional~\cite{hughes}. In the
present case, the generalized paraboson algebra related to $R$
can allow an isolated number of unitary finite-dimensional
representations (apart from the infinite-dimensional ones). In
the case $\deg(f)=2$ there can be one such isolated example with
dimension~2. We have come across other such examples for
$\deg(f)>2$, e.g.\ when $f(x)=x^3+x^2-2x+2$, the module $L(0)$
is 3-dimensional and it is also a unitary representation with
matrix elements given (after normalization) explicitly by
\beq
A=\left(\begin{array}{ccc} 0&\sqrt{2}&0\\0&0&\sqrt{2}\\
 0&0&0\end{array}\right),\qquad
B=\left(\begin{array}{ccc} 0&0&0\\ \sqrt{2}&0&0\\
0&\sqrt{2}&0\end{array}\right),\qquad
H=\left(\begin{array}{ccc} 0&0&0\\0&-1&0\\0&0&-2\end{array}\right).
\eeq

\section{Deformed oscillator algebras}

A generalized deformed oscillator algebra is defined in terms of
three generators $a,\ad,N$ subject to the relations
\beq
[N,\ad]=\ad,\qquad [N,a]=-a,\qquad \ad a=\tilde F(N),\qquad
a\ad=\tilde F(N+1).
\label{aa}
\eeq
There is a close relationship with the algebra generated by the
same generators but subject to
\beq
[N,\ad]=\ad,\qquad [N,a]=-a,\qquad a\ad+\ad a=\tilde f(N),
\label{aaaa}
\eeq
where $\tilde f(t)=\tilde F(t)+\tilde F(t+1)$. From Lemma~\ref{lem2}
we know that $\tilde f$ and $\tilde F$ determine each other uniquely.
\def\bd{b^\dagger}
As is well known, the operators $a, \ad$ and $N$ can under
certain restrictions be realized
in terms of ordinary boson operators $b=(x+ip)/\sqrt{2}$ and
$\bd=(x-ip)/\sqrt{2}$ as follows
\beq
N=\bd b,\qquad a=\sqrt{\tilde F(N+1) \over N+1}\, b,\qquad
\ad = \bd \sqrt{\tilde F(N+1) \over N+1} .
\eeq
For the ordinary boson operator with $[b,\bd]=1$ there exists
the Fock representation with states $\ps_n$
($n=0,1,2,\ldots$), where $b\ps_n=\sqrt{n}\ps_{n-1}$ and $\bd
\ps_n=\sqrt{n+1}\ps_{n+1}$. This becomes a representation of the
deformed algebra if $\tilde F(n)>0$ for $n=1,2,\ldots$ -- this
could be called the Fock representation of the generalized
oscillator algebra. Note that its basis vectors are the states
of the ordinary Fock space. If we consider
\beq
{\cal H}={1\over 2} (a\ad + \ad a)={1\over 2} \tilde f(N)
\eeq
as the Hamiltonian of a system associated with the generalized
oscillator algebra, then its
energy spectrum is given by
\beq
E_n = {1\over 2} \tilde f(n),\qquad n=0,1,2,\ldots.
\eeq
The interesting question is whether there exist physical systems
with such an energy spectrum. In fact, in certain cases one can
find physically realizable systems using the above boson
realization. One just has to substitute in $\tilde f(N)$ the
boson realization for $N$, namely
\beq
N=\bd b ={1\over 2} (p^2+x^2-1),
\eeq
and see whether ${1\over 2}\tilde f(N)$ can correspond to the
Hamiltonian of any physically realizable system.

For example, let $\tilde f(t)=2(t+1)^2$. Then
\bea
{\cal H}&=&(N+1)^2=(p^2+x^2+1)^2/4={1\over 2}p^2+V(x,p),\\
\hbox{with }V(x,p)&=& {1\over 2}x^2+{1\over
4}(x^4+p^4+x^2p^2+p^2x^2) +{1\over 4}.
\eea
This is a physical system with a potential $V(x,p)$ depending on both the
position and the velocity, and its energy eigenvalues are given
by $E_n=(n+1)^2$, $n=0,1,\ldots$. Furthermore, the
eigenfunctions of this system would be the same as the
eigenfunctions of the ordinary harmonic oscillator of unit mass
and unit circular frequency. Hence one can distinguish the
ordinary harmonic oscillator from the above system not by their
eigenfunctions but only by the quantum jumps! Another example of
$\tilde f(t)$ is available in~\cite{brodimas}.

To see the connection between the above Fock representation and
the unitary modules discussed in the previous section, one can
make for the above example the following identification~:
\beq
A=a, \qquad B=\ad,\qquad H=-N-3/2.
\eeq
Then we find the usual structure (\ref{defrel}) for $R$ with
$f(t)=2t^2+2t+1/2$. Hence, $f$ is in the standard form (\ref{deg2})
with $c=1/2$. The Fock representation with $N$-lowest weight $0$
corresponds to the module $L(\la)$ with $H$-highest weight
$\la=-3/2$. For $c=1/2$ the module $L(-3/2)$ is indeed unitary.
All the other unitary modules, including the 2-dimensional one,
will be consistent with the relations $a\ad+\ad a=\tilde
f(N)=2(N+1)^2$ and thus all correspond to unitary
representations of (\ref{aaaa}). For example, the 2-dimensional module
(\ref{2dim}) would yield (after normalizing)~:
\beq
a=\left(\begin{array}{cc} 0&\sqrt{1/2}\\0&0\end{array}\right),\qquad
\ad=\left(\begin{array}{cc} 0&0\\\sqrt{1/2}&0\end{array}\right),\qquad
N=\left(\begin{array}{cc} -3/2&0\\0&-1/2\end{array}\right).
\eeq
These satisfy indeed $[N,a]=-a$, $[N,\ad]=\ad$ and $a\ad+\ad
a=2(N+1)^2$. But among all the unitary representations, only in
$L(-3/2)$ the extra relations $\ad a=\tilde F(N)$ and
$a\ad=\tilde F(N+1)$ are valid. So $L(-3/2)$ is the only unitary
representation of (\ref{aaaa}) that is also a unitary
representation of (\ref{aa}). This is the property that makes the Fock
representation different from the others.

This difference between (\ref{aa}) and (\ref{aaaa}) can be
considered as the generalization of the difference between
ordinary bosons and parabosons. For ordinary bosons there is
only the relation $b\bd-\bd b=1$, and only the Fock space. For
parabosons, the relations are
\beq
[\{B^-,B^+\},B^\pm]=\pm 2B^\pm,
\eeq
and now all unitary representations of $osp(1/2)$ are consistent
with these relations. Among all these, the
one corresponding to the Fock space is the only representation where the
extra relation $B^-B^+-B^+B^-=1$ is valid. In this sense
(\ref{aa}) could be called a generalized boson algebra, and
(\ref{aaaa}) could be called a generalized paraboson algebra.

\section*{Acknowledgements}
It is a pleasure to thank Prof.\ T.D.\ Palev for stimulating
discussions. One of us (R.J.) wishes to thank Prof.\
Guido Vanden Berghe for the kind hospitality at the Department
of Applied Mathematics and Computer Science, University of Ghent.
This research was partly supported by the E.E.C. (contract No.
CI1*-CT92-0101).

\end{document}